\newcommand{\bm}[1]{\mbox{\boldmath$#1$}}
\documentstyle[multicol,prb,aps,epsbox]{revtex}

\input epsf

\title{Binder Parameter of a Heisenberg Spin-Glass Model 
in Four Dimensions }

\author{ Takayuki {\sc Shirakura} and Fumitaka {\sc Matsubara}$^1$}

\address{Faculty of Humanities and Social Sciences, Iwate University, 
Morioka 020-8550, Japan \\
$^1$Department of Applied Physics, Tohoku University, Sendai 980-8579,
Japan}

\date{ \today }

\begin{document}

\maketitle

\begin{abstract}

We studied the phase transition of the $\pm J$ Heisenberg model 
with and without a random anisotropy on four 
dimensional lattice $L\times L\times L\times (L+1)$ $(L\leq 9)$. 
We showed that the Binder parameters $g(L,T)$'s for different sizes 
do not cross even when the anisotropy is present. 
On the contrary, when a strong anisotropy exists, $g(L,T)$ exhibits 
a steep negative dip near the spin-glass phase transition temperature 
$T_{\rm SG}$ similarly to the $p-$state infinite-range Potts glass 
model with $p \geq 3$, in which the one-step replica-symmetry-breaking 
(RSB) occurs. We speculated that a one-step RSB-like state occurs 
below $T_{\rm SG}$, which breaks the usual crossing behavior of $g(L,T)$.


\end{abstract}

%%%%%%%%%%%%%%%%%%%%%%%%%%%
\begin{multicols}{2}
%%%%%%%%%%%%%%%%%%%%%%%%%%

%\pacs{75.50.Lk,02.70.Lq,05.50.+q}

\newpage
%%%%%%%%%%%%%%%%%%%%%%%%%%%%%%%%%%%%%%%%%%%%%%%%%%%%%%%%%%%%%%%%%%%%%%%%%%%%%

Recently, the low temperature phase of the XY and Heisenberg spin-glass (SG) 
models has been attacked a great interest. 
It is known that vector SG models have a chiral symmetry in addition to 
the continuous symmetry\cite{Villain}. 
Consequently, these models may have both SG order 
and chiral glass (CG) order. 
Kawamura and coworkers claimed that decoupling of the spin and chiral 
valuables occurs at long distances\cite{Kawamura1,Kawamura2}, and that 
the CG order is realized at low temperatures in three dimensions ($d=3$), 
but the SG order is not\cite{Kawamura3,Kawamura4,Kawamura5,Kawamura6}. 
Although the existence of the CG order has been accepted, 
controversy exists on the SG order. 
Maucourt and Grempel\cite{Maucourt} studied the domain wall energy $W(L)$ 
of the $\pm J$ XY model on finite $d=3$ lattices of $L^3$ 
at absolute zero temperature ($T = 0$) and speculated that $W(L)$ 
may increase with the linear size $L$. The same speculation was also given 
by Kosterlitz and Akino\cite{KA}. 
The present authors examined the stiffness of the $\pm J$ Heisenberg model 
at $T = 0$\cite{Matsu1} and $T \neq 0$\cite{Matsu2}, and suggested 
that the stiffness exponent $\theta$ in $d=3$ has a positive 
value for $T/J \lesssim 0.19$.
They also performed a Monte-Carlo (MC) simulation of the same model 
and found that the SG susceptibility $\chi_{\rm SG}$ exhibits 
a divergent behavior toward the temperature of $T/J \sim 0.18$\cite{Shira}.
These facts strongly suggests that the SG order also occurs in the XY 
and Heisenberg models in $d=3$. 
This view of the SG order was supported in recent studies of 
the aging effect of the spin autocorrelation function 
$\langle\bm{S}_i(0)\bm{S}_i(t)\rangle$\cite{Matsu3} and 
the non-equilibrium relaxation of $\chi_{\rm SG}$\cite{Nakamura}. 
However, the Binder parameters $g(L,T)$'s for different $L$ 
neither cross nor come together in both the XY\cite{Kawamura6} 
and Heisenberg\cite{Kawamura5} SG models. 

% in the positive region of $g(L,T)$. 
This fact raises an objection against the occurrence of the SG order, 
because it is believed that the most reliable evidence of the occurrence 
of the SG phase transition is the crossing of $g(L,T)$'s at the same 
temperature of $T_{\rm SG}$\cite{Bhatt,Kawashima,Marinari1}. 
In fact, it was reported that the crossing of $g(L,T)$'s occurs in 
both the XY\cite{Jain} and Heisenberg\cite{Coluzzi} SG models in 
four dimensions ($d=4$). 
Thus we are faced by a serious problem that {\it we have 
gotten the opposite views of the SG order from the studies of 
the different quantities}. 

%%%%%%%%%%%%%%%%%%%%%%%%%%%%%
The problem may arise from a poor knowledge of the property of 
$g(L,T)$ of the vector SG model. 
In the non-frustrated system, the crossing of $g(L,T)$'s occurs at 
$T_{\rm C}$ with some positive value of $\tilde{g}$ ($\equiv g(L,T_{\rm C})$). 
This property results from the fact that, in the thermodynamic limit, 
$g(\infty,T) = 0$ for $T > T_{\rm C}$ and $g(\infty,T) = 1$ for 
$T < T_{\rm C}$. 
However, the same will not always be true in the SG model, because 
$g(\infty,T)$ for $T < T_{\rm SG}$ will take different values 
due to the occurrence of the replica symmetry breaking (RBS). 
In fact, it was revealed by Hukushima and Kawamura\cite{Hukushima1} that 
$g(\infty,T_{\rm SG}^-)$ of the infinite-range $p$-state Potts glass 
model\cite{Sompolinsky} 
takes different values depending on the state number $p$, where 
$T^-=\lim_{\epsilon \rightarrow 0}(T-\epsilon)$. 
Therefore it is crucially important to reveal the property 
of $g(L,T)$ in such vector SG models in which the SG phase transition 
occurs at $T_{\rm SG} \neq 0$. 

%%%%%%%%%%%%%%%%%%%%%%%%%%%%%
In this paper, we report that $g(L,T)$ of the Heisenberg SG model in $d=4$ 
exhibits a behavior quite different from that of the Ising SG model.
We reexamined $g(L,T)$ of the $\pm J$ Heisenberg models in $d=4$ 
with and without a random anisotropy of magnitude $D$. 
In the case of $D = 0$, when $L$ is increased, 
the increment of $g(L,T)$ with decreasing temperature becomes 
less steep at low temperatures, and the crossing of $g(L,T)$'s which was 
suggested for small $L$\cite{Coluzzi} is released. 
When $D \neq 0$, $g(L,T)$ for 
each $L$ decreases, particularly around $T_{\rm SG}$, and this 
property is enhanced when $D$ is increased, where $T_{\rm SG}$ is the
SG transition temperature estimated from the scaling plot of 
$\chi_{\rm SG}$. 
In particular, $g(L,T)$ for large $D$ exhibits  
a negative dip near $T_{\rm SG}$ which deepens with increasing $L$. 
These facts are quite interesting, because the anisotropy 
would stabilize the SG order\cite{BrayMY,Iyota2}. 
We believe, hence, that {\it the absence of the crossing of $g(L,T)$'s 
for finite $L$ says nothing about the presence of the SG order 
in this model.} 
We speculate that, even for $D=0$, $g(\infty,T_{\rm SG}^-)$ takes 
a small positive value (or a negative value) due to the occurrence 
of a one-step RSB-like state.
We hope our findings will help to understand the low 
temperature phase of the Heisenberg SG model in $d=3$.

%%%%%%%%%%%%%%%%%%%%%%%%%%%%%%%%%%%%%%%%%%%%%%%%%%%%%%%%%\newpage
We start with the anisotropic $\pm J$ Heisenberg model on a hyper cubic 
lattice of $L \times L \times L \times (L+1) ( \equiv N) $ with skew 
boundary conditions along three $L$ directions and a periodic boundary
condition along the $(L+1)$ direction.
The Hamiltonian is described by 
\begin{eqnarray} 
     H = - \sum_{\langle ij \rangle}[J_{ij}\bm{S}_{i}\bm{S}_{j}
		 + \sum_{\alpha\neq\beta}D_{ij}^{\alpha\beta}S_i^\alpha S_j^\beta], 
\end{eqnarray} 
where $\bm{S}_{i}$ is the Heisenberg spin of $|\bm{S}_i| = 1$ 
and $S_i^{\alpha}$ is its $\alpha$-component $(\alpha = x, y, z)$, 
and $\langle ij \rangle$ runs over all nearest-neighbor pairs. 
The exchange interaction $J_{ij}$ takes on either $+J$ or $-J$ with 
the same probability of 1/2. We assume that the anisotropy comes from
pseudo-dipolar couplings and impose the restriction
$D_{ij}^{\alpha\beta} = D_{ji}^{\alpha\beta} = D_{ij}^{\beta\alpha}$.
We further assume that $D_{ij}^{\alpha\beta}$ are uniform random values 
between $-D$ and $D$. 
We note that the role of the anisotropy is to break the rotational 
symmetry of the model and to stabilize the SG order. 

%%%%%%%%%%%%%%%%%%%%%%%%%%%%%%%%%%%%%%%%%%%%%%%%%%%%%%%%%%%%%%%%%%
We performed a MC simulation of the two-replica systems of $\{\bm{S}_i\}$ 
and $\{\bm{\tilde{S}_i}\}$ using an exchange MC algorithm\cite{Hukushima2}.
We calculated the order-parameter probability distribution $P_L(q)$ of 
\begin{eqnarray}
	P_L(q) = [\langle\delta (q - Q)\rangle],
\end{eqnarray}
where $\langle \cdots \rangle$ and [$\cdots$] mean the thermal average 
and the bond distribution average, respectively. 
Here $Q$ is the spin overlap defined by 
\begin{eqnarray}
	Q = \sqrt{\frac{1}{3}\sum_{\alpha,\beta} (q^{\alpha\beta})^2},
\end{eqnarray}
with $q^{\alpha\beta} \equiv \frac{1}{N} \sum_{i=1}^N S_i^\alpha 
\tilde{S}_i^\beta$.
Using $P_L(q)$, we obtained the SG susceptibility $\chi_{\rm SG}$ 
and the Binder parameter $g(L,T)$ which are defined by 
\begin{eqnarray} 
   \chi_{\rm SG} &=& 3N[\langle q^2\rangle],\\  
	g(L,T) &=& \frac{1}{2} (11 - 9 \frac{[\langle q^4\rangle]}
			{[\langle q^2\rangle]^2}),
\end{eqnarray}
where $[\langle q^n \rangle] = \int q^n P_L(q) dq$. 
To examine whether the SG order occurs or not, we also calculated 
quantities $A(L,T)$ and $G(L,T)$ that measure the order-parameter 
fluctuations (OPF): 
\begin{eqnarray}
	A(L,T) &=& \frac{[\langle q^2\rangle^2] - [\langle q^2\rangle]^2}
               {[\langle q^2\rangle]^2},   \\
	G(L,T) &=& \frac{[\langle q^2\rangle^2] 
	- [\langle q^2\rangle]^2}{[\langle q^4\rangle] - [\langle q^2\rangle]^2}.
\end{eqnarray}
Each of these quantities will exhibits a crossing behavior at $T_{\rm SG}$ 
when the SG transition occurs\cite{Marinari2,Picco}. 
We performed the MC simulation of the model (1) with various values 
of $D = 0, 0.1J, 0.2J, 0.5J$ and $1.0J$. 
The linear sizes of the lattice studied here are $L = 3 \sim 9$. 
Equilibration was checked by monitoring the stability of the results 
against at least two-times longer runs. 
The numbers of the samples were 480 for $L = 3$, 288 for $L = 5$, 
96 for $L = 7$, and 48 for $L = 9$.

%%%%%%%%%%%%%%%%%%%%%%%%%%%%%%%%%%%%%%%%%%%%%%%%%%%%%%%%%%%%%%%%%%%
%%%%%%%%%%%%%%%%%%%%%%%%%%%%%%%%%%%%%%%%%%%%%%%%%%%%%%%%%%%%%%%%%%%
%
\begin{figure}
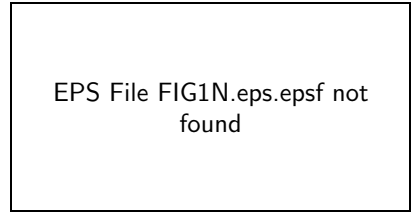

\psbox[scale=0.45]{FIG1N.eps}
%\epsfbox{FIG1.eps}
\caption{
(a) Temperature dependences of the spin-glass susceptibility $\chi_{\rm SG}$ 
of the $\pm J$ Heisenberg model in $d=4$ for different sizes of the lattice. 
Open symbols represent those for $D=0$ and closed ones those for $D=0.5J$. 
(b) An example of the scaling plot of $\chi_{\rm SG}$ for $D = 0$.
}
\label{fig:1}
\end{figure}
In Fig. 1(a), we show results of $\chi_{SG}$ in $D = 0$ and $D = 0.5J$. 
In both the cases, $\chi_{SG}$ for larger $L$ increases rapidly 
as the temperature is decreased. 
The same was true for different values of $D$. 
The finite size scaling analysis in each $D$ suggested the divergence of 
$\chi_{\rm SG}$  for $L \rightarrow \infty$ at a finite, non-zero 
temperature. An example of the scaling plot in the case of $D = 0$ 
is presented in Fig. 1(b). 
Hereafter, we tentatively call this temperature the SG transition temperature 
and denote $T_{\rm SG}$. 
Note that $T_{\rm SG}$ increases with increasing $D$.
These results are compatible with a common belief that the SG order 
occurs in $d=4$ even for $D = 0$, and that the SG order is stabilized by 
the anisotropy\cite{BrayMY,Iyota2}. 

%%%%%%%%%%%%%%%%%%%%%%%%%%%%%%%%%%%%%%%%%%%%%%%%%%%%%%%%%%%%%%%%%%%%
%
%
Now we show $g(L,T)$ in different $D$'s in Figs. 2(a) - 2(c), 
When $D = 0$, $g(L,T)$'s for $L=3$ and 5 come close to each other 
near $T_{\rm SG}$. But they do not cross, because the increment of 
$g(L,T)$ for $L =5$ is suppressed below $T_{\rm SG}$.  
This suppression is not released for larger $L$. 
Therefore we believe that, in contrary to the previous 
suggestion\cite{Coluzzi}, $g(L,T)$'s for large $L$ do not cross 
but converge on some non-zero value $\tilde{g}$
at $T=T_{\rm SG}$.\cite{Comment0}. 
When $D \neq 0$, the suppression is enhanced more. 
In particular, for large $D$, $g(L,T)$ exhibits a dip near $T_{\rm SG}$ 
which deepens as $L$ is increased. 
This result is also incompatible with our naive expectation that, 
as $D$ is increased, $g(L,T)$'s would tend to cross with some positive 
$\tilde{g}$, because the anisotropy will stabilize the SG order. 
%%%%%%%%%%%%%%%%%%%%%%%%%%%%%%%%%%%%%%%%%%%%%%%%%%%%%%%%%%%%%%%%%%%
%
\begin{figure}
\psbox[scale=0.42]{FIG2N.eps}
\caption{
Temperature dependences of the Binder parameter $g(L,T)$ of 
the $\pm J$ Heisenberg model in $d=4$ for different magnitude 
of the anisotropy $D$ and for different sizes of the lattice; 
(a) $D=0$, (b) $D=0.1J$, and (c) $D=0.5J$ (open symbols) and 
$D=1.0J$ (closed symbols). Arrows indicate the SG transition 
temperature $T_{\rm SG}$ 
estimated from the scaling plots of the SG susceptibility. 
}
\label{fig:2}
\end{figure}
We next examined $G(L,T)$ which would exhibit a crossing behavior 
at $T_{\rm SG}$ even when $g(L,T)$ did not exhibit the crossing 
behavior\cite{Hukushima1,Picco}. 
Here, in Figs. 3(a) and 3(b), we show $G(L,T)$'s for different $L$ in the 
cases of $D = 0$ and $D = 0.5J$, respectively. 
When $D = 0$ (and also $D \leq 0.2J$), $G(L,T)$'s for large $L (\geq 5)$ 
seem to come together near $T_{\rm SG}$.  
This property becomes more prominent in large $D$, 
and the data for $L = 3$ joins. 
A similar crossing has also been found in the other quantity $A(L,T)$. 
\begin{figure}
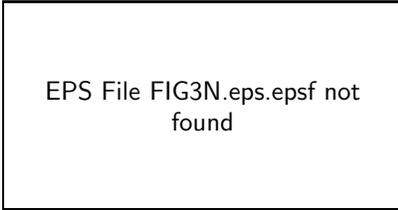

\psbox[scale=0.44]{FIG3N.eps}
%\epsfbox{FIG3N.eps}
\caption{
Temperature dependences of the order-parameter fluctuations(OPF) $G(L,T)$ 
of the $\pm J$ Heisenberg model in $d=4$ for different sizes of the lattice 
and for different magnitude of the anisotropy; (a) $D=0$ and (b) $D=0.5J$. 
Arrows indicate the SG transition temperature $T_{\rm SG}$
estimated from the scaling plots of the SG susceptibility. 
}
\label{fig:3}
\end{figure}

%%%%%%%%%%%%%%%%%%%%%%%%%%%%%%%%%%%%%%%%%%%%%%%%%%%%%%%%%%%%%%%%%%%
We have calculated four quantities $\chi_{\rm SG}$, $g(L,T)$, 
$G(L,T)$ and $A(L,T)$ to examine the SG phase transition. 
All the quantities except for $g(L,T)$ suggested the occurrence of 
the SG order below $T_{\rm SG}$. 
Considering the facts that $G(L,T)$ and $A(L,T)$ can be used to determine 
the value of $T_{\rm SG}$ of such SG models in which $T_{\rm SG}$ is hardly 
determined by the usual crossing behavior of $g(L,T)$\cite{Hukushima1,Picco}, 
and that the $L$-dependence of $\chi_{\rm SG}$ clearly suggests the 
divergence of the spin correlation length at $T_{\rm SG}$, 
it is natural to conclude that the SG order occurs below $T_{\rm SG}$. 
Therefore, the absence of the crossing of $g(L,T)$'s for finite $L$ 
would say nothing about the SG phase transition in this model.

%%%%%%%%%%%%%%%%%%%%%%%%%%%%%%%%%%%%%%%%%%%%%%%%%%%%%%%%%%%%%%%%%%%%%%%%%%
What occurs in $g(L,T)$ at $T_{\rm SG}$ for $L \rightarrow \infty$? 
The occurrence of a dip of $g(L,T)$  near $T_{\rm SG}$ for large $D$ 
($D = 0.5J$ and $1.0J$) is suggestive. 
The dip deepens as $L$ is increased implying a negative divergence of 
$g(\infty,T_{\rm SG}^-)$. 
It is known that, in the infinite-range $p$-state Potts glass, 
the value of $g(\infty,T_{\rm PG}^-)$ changes continuously from 1 for 
$p = 2$ (Ising model) to $-\infty$ for $p = 4$ due to the occurrence 
of the one-step RSB for $p > 2$\cite{Hukushima1,Sompolinsky}. 
Therefore we suggest that $g(\infty,T)$ takes $-\infty$ or some very large 
negative value at $T_{\rm SG}^-$ due to the occurrence of a one-step 
RSB-like state. 
In fact, as shown in Fig. 4, the order parameter distribution function 
$P_L(q)$ for $D = 1.0J$ exhibits a double peak at low 
temperatures\cite{Comment1}. 

%%%%%%%%%%%%%%%%%%%%%%%%%%%%%%%%%%%%%%%%%%%%%%%%%%%%%%%%%%%%%%%%%%%%%%%
%
\begin{figure}
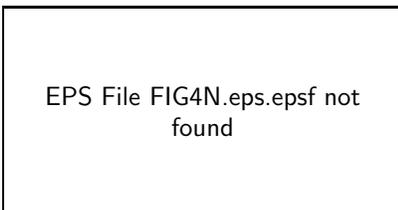

\psbox[scale=0.40]{FIG4N.eps}
%\epsfbox{FIG4N.eps}
\caption{
The order-parameter probability distribution $P_L(q)$ 
of the $\pm J$ Heisenberg model in $d=4$ for $D=1.0J$ 
at a low temperature of $T = 0.6J ( < T_{\rm SG} \sim 0.95J)$.
} 
\label{fig:4}
\end{figure}

On the other hand, in the case of small $D$, no distinct dip but a bending of 
$g(L,T)$ was seen around $T_{\rm SG}$.  
We may also explain this behavior based on a plausible assumption that 
$g(\infty,T_{\rm SG}^-)$ takes a negative value or a small positive 
value due to the occurrence of the one-step RSB-like state. 
When $D$ = 0, the rotational symmetry recovers and $g(\infty,T)$ might 
exhibit some different property. 
We think, however, that $g(\infty,T_{\rm SG}^-)$ also takes 
a small positive value (or a negative value), because our data of 
Figs. 2(a) - 2(c) imply the presence of no gap between the cases of 
$D = 0$ and $D \neq 0$. 
Of course, we could not rule out the possibility that there exists some 
threshold $D^*$, including $D^*=0$, below which the nature of $g(\infty,T)$ 
changes qualitatively. 

%%%%%%%%%%%%%%%%%%%%%%%%%%%%%%%%%%%%%%%%%%%%%%%%%%%%%%%%%%%%%%
In summary, we reexamined the spin-glass phase transition of  
the $\pm J$ Heisenberg model in four dimensions ($d=4$) and gave 
a confirmation that the SG transition really occurs even when the 
anisotropy is absent $D = 0$. However, its transition temperature 
$T_{\rm SG}$ could not be determined from the usual crossing behavior 
of the Binder parameter $g(L,T)$\cite{Coluzzi}, but from the crossing 
of $G(L,T)$'s (and also $A(L,T)$'s),  
as well as from the divergence 
of the SG susceptibility $\chi_{\rm SG}$. 
This fact was quite interesting, because it threw a doubt on the 
common belief that the most reliable evidence of the occurrence of the 
SG order is the crossing of $g(L,T)$'s. 
We speculated that the low temperature phase of the vector SG 
model is characterized by one-step RSB-like 
state and the crossing of $g(L,T)$'s for finite $L$ is absent.

\bigskip

The authors would like to thank Mr. S. Endoh and Dr. T. Nakamura 
for their valuable discussions.

%%%%%%%%%%%%%%%%%%%%%%%%%%%%%%%%%%%%%%%%%%%%%%%%%%%%%%%%%%%%%%%%%%%

%%%%%%%%%%%%%%%%%%%%%%%%%%%
\end{multicols}
%%%%%%%%%%%%%%%%%%%%%%%%%%


\begin{references}
\bibitem{Comment0}
		The crossing behavior of $g(L,T)$ at $T = T_{\rm SG}$ comes from 
		the combination of (A) the scaling hypothesis 
		and (B) the assuption $g(\infty,T) = 0$ for $T > 
		T_{\rm SG}$ and $g(\infty,T) = 1$ for $T < T_{\rm SG}$.  Therefore 
		when the SG phase transition occurs, even if the assumption (B) 
		is not satisfied, $g(L,T)$ will converge on some non-zero value 
		at $T = T_{\rm SG}$. 
		We think a large finite size effect masks this property.
\bibitem{Comment1}
Note that a double peak was found in the chirality order parameter 
distribution $P_L^{(\rm c)}(q)$ of the Heisenberg\cite{Kawamura5} 
and XY\cite{Kawamura6} SG models with $D = 0$ in $d=3$ 
in which the one-step RSB was suggested to occur.
\end{references}
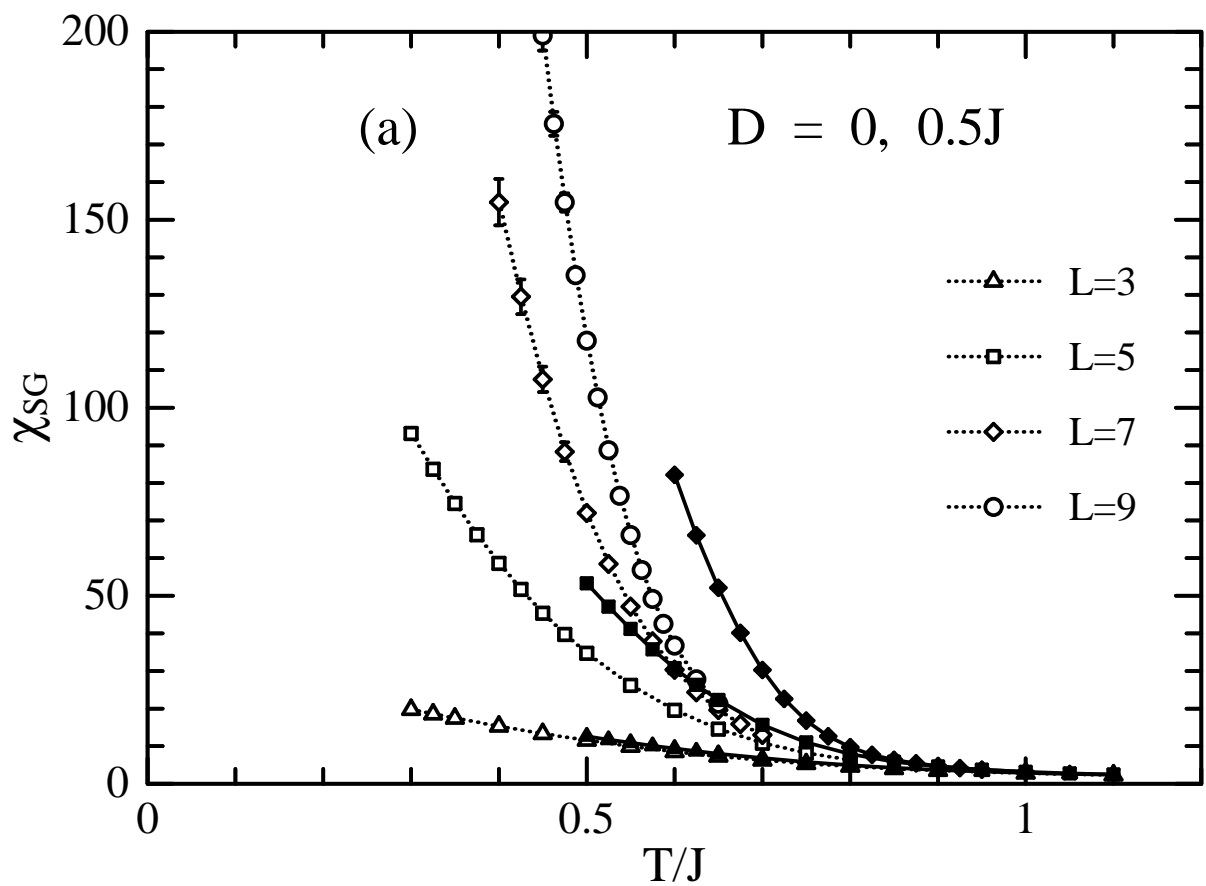
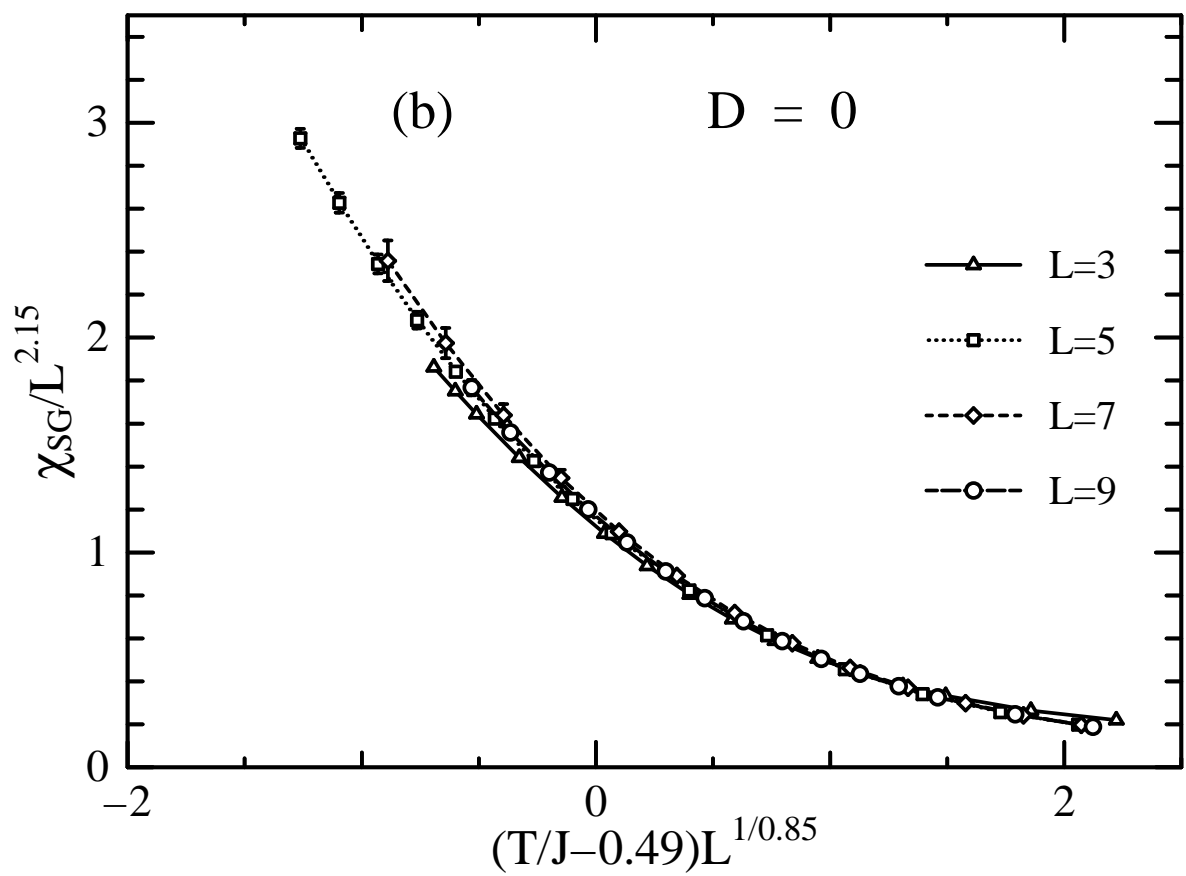
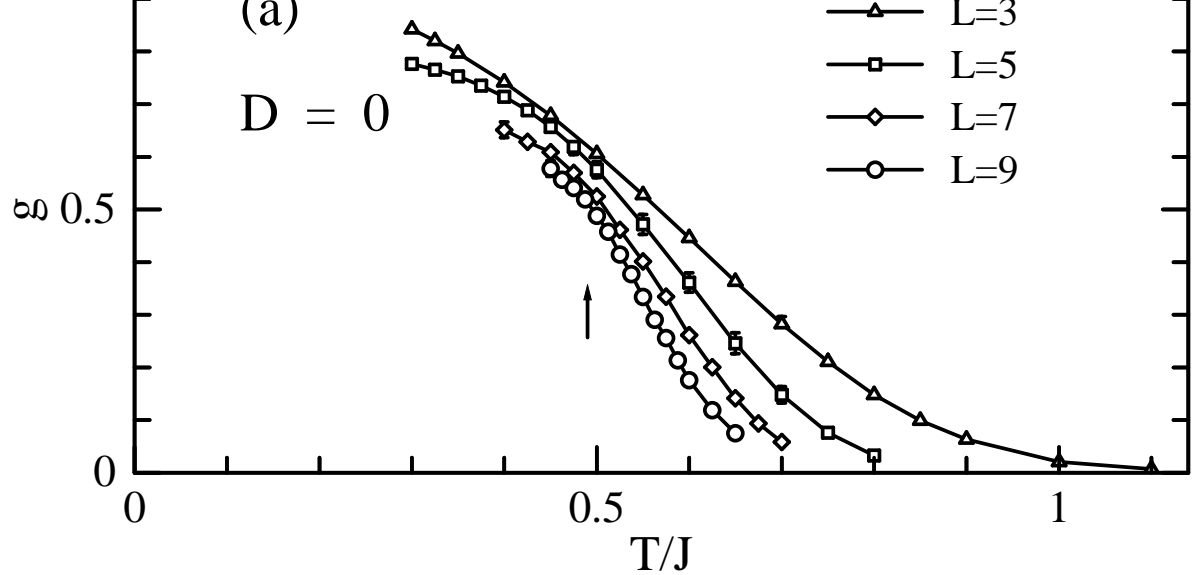
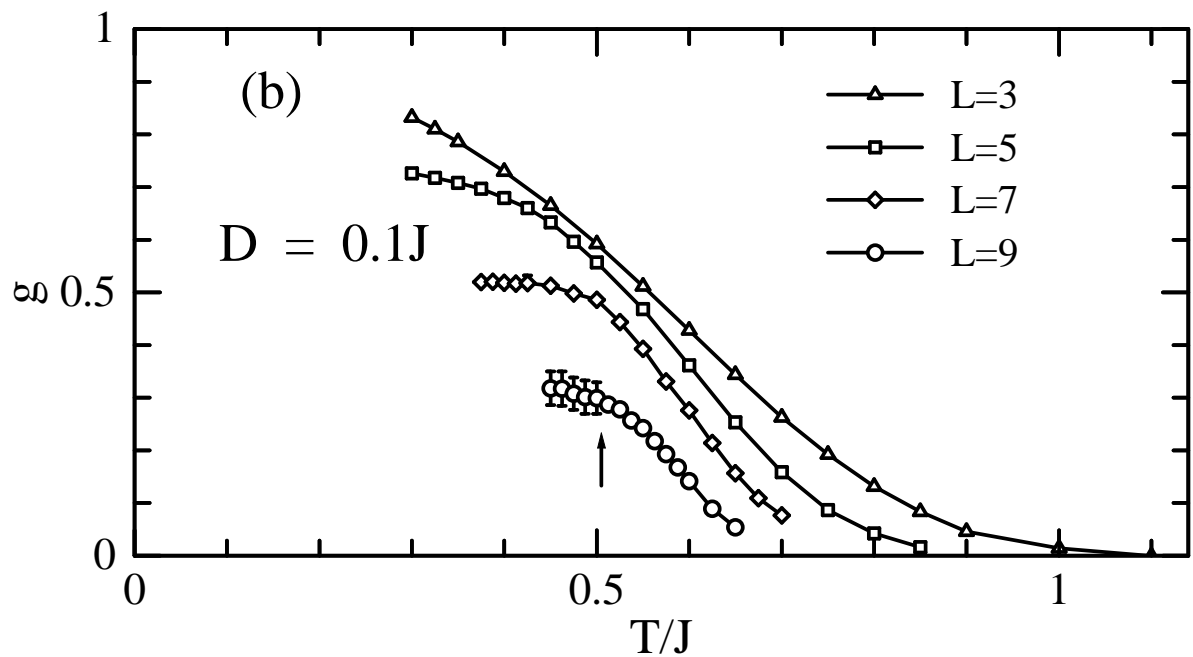
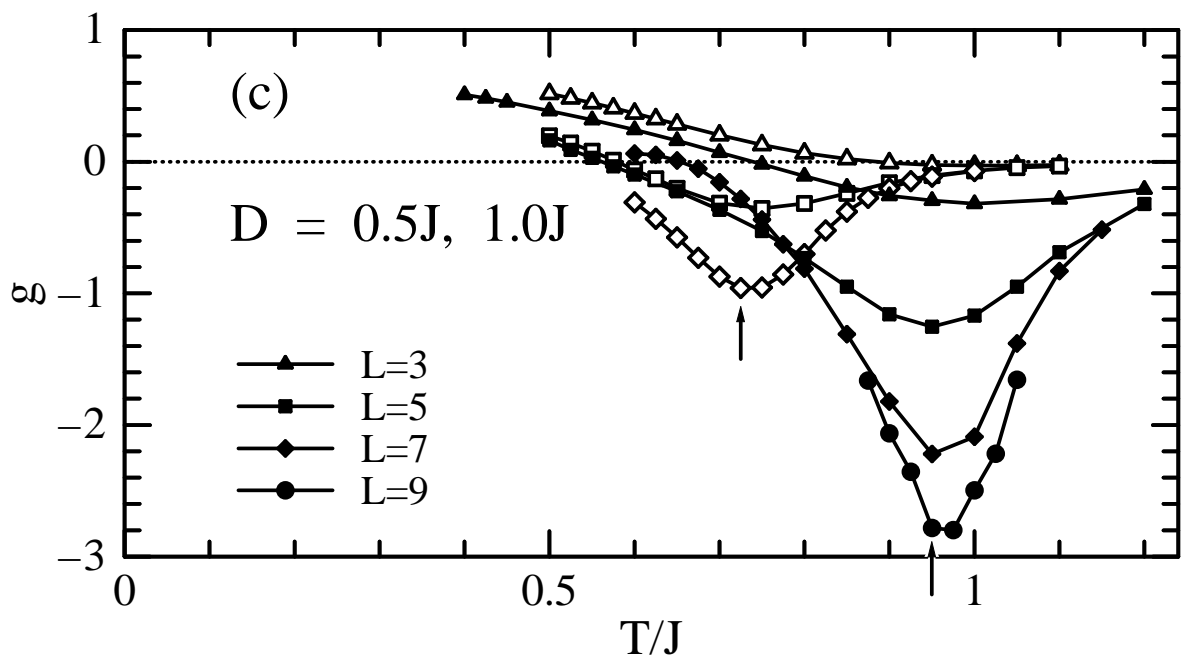
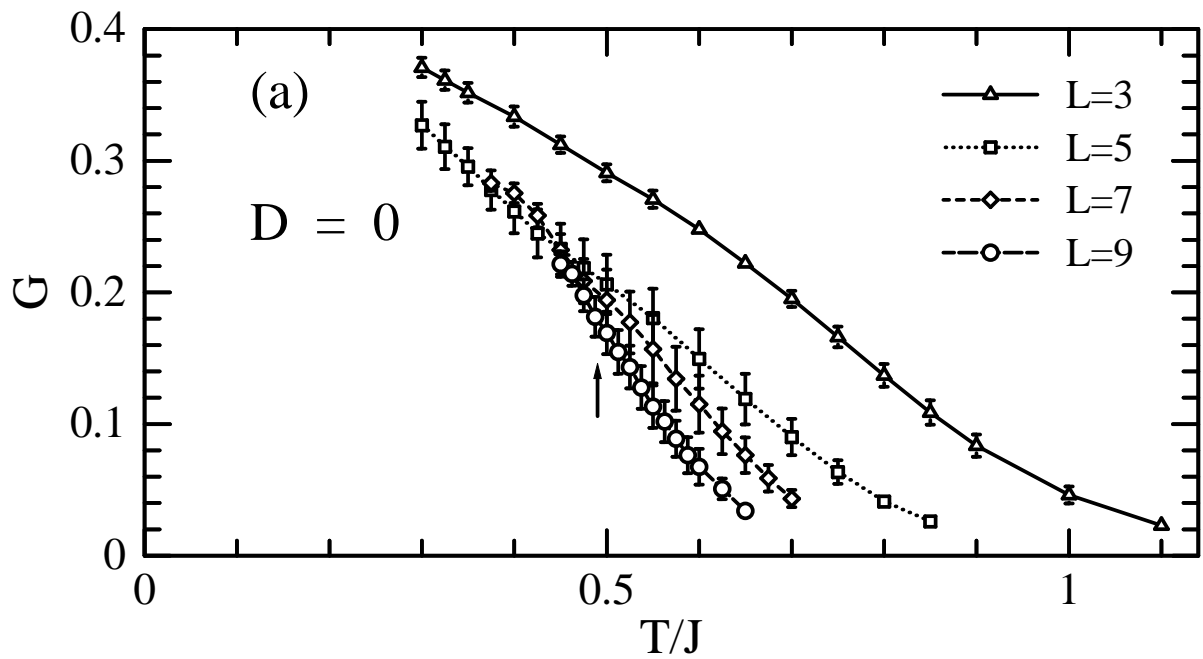
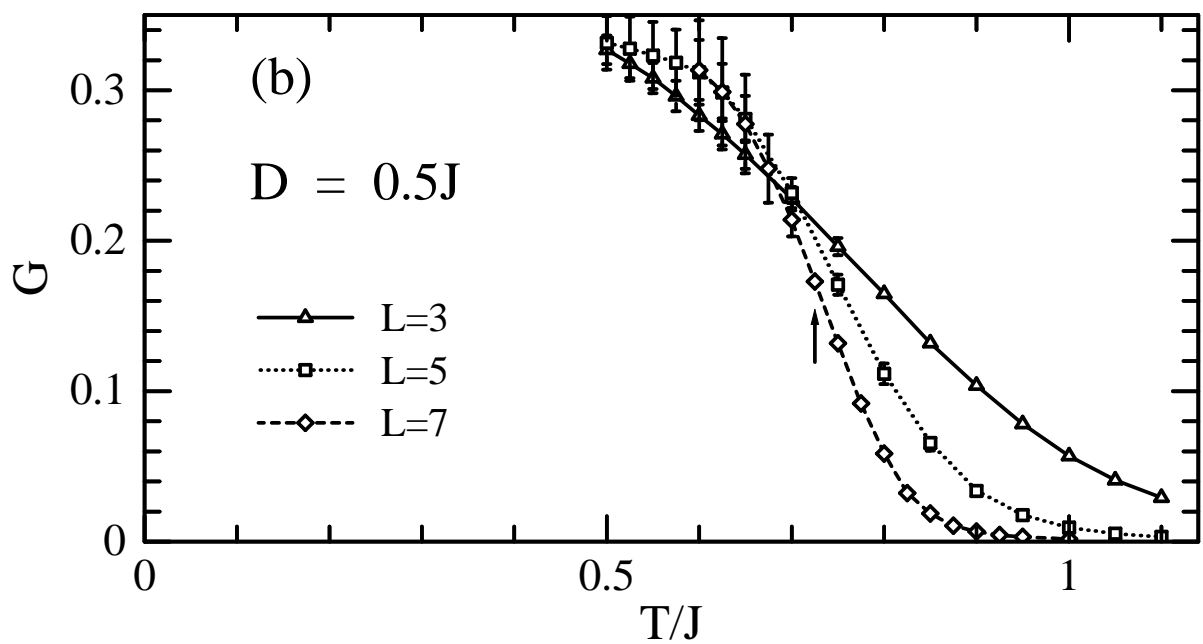
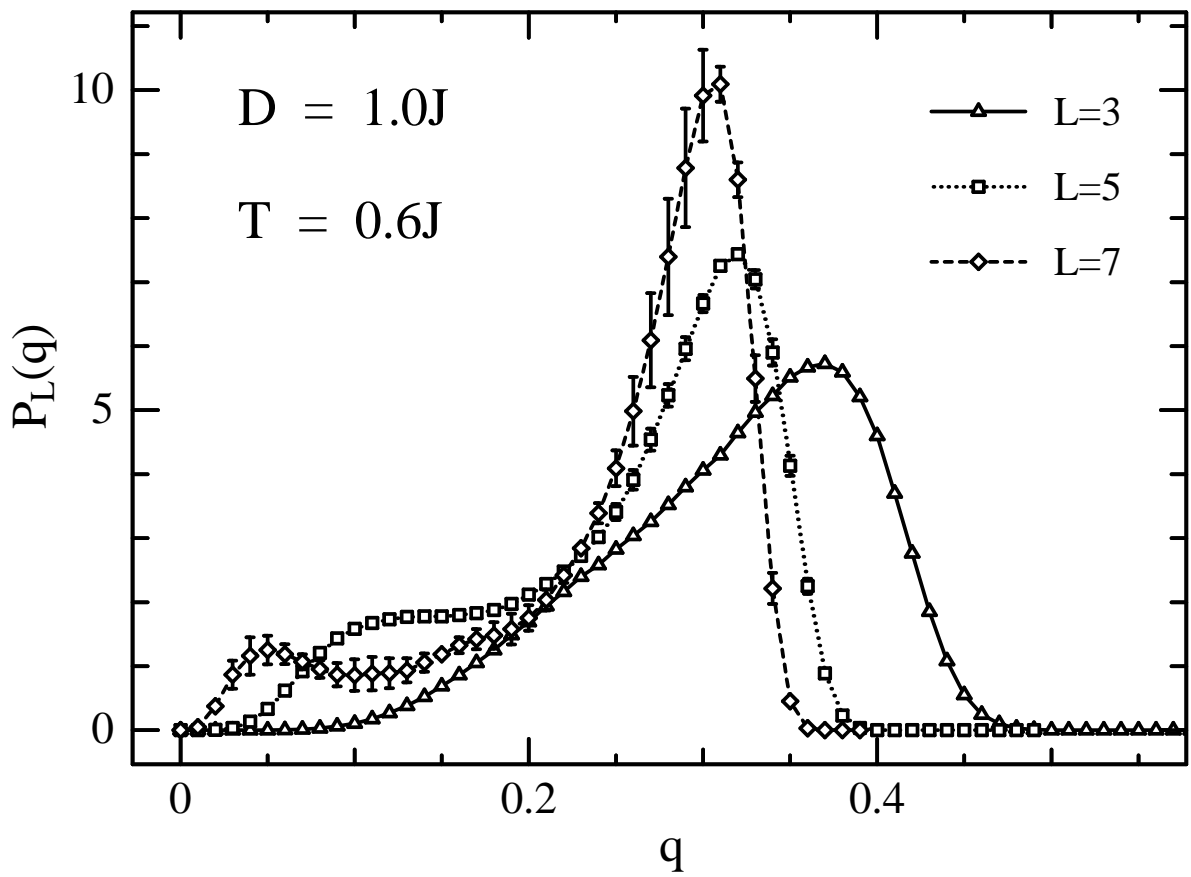
\end{document}